# Exploring effective charge in electromigration using machine learning

Yu-chen Liu[1,2], Benjamin Afflerbach[2], Ryan Jacobs[2], Shih-kang Lin[1,3,4], and Dane Morgan[2,*]

[1] Department of Materials Science and Engineering, National Cheng Kung University, Tainan city 70101, TAIWAN; [2] Department of Materials Science and Engineering, University of Wisconsin-Madison, Madison, USA; [3] Center for Micro/Nano Science and Technology, National Cheng Kung University, Tainan city 70101, TAIWAN; [4] Hierarchical Green-Energy Materials (Hi-GEM) Research Center, National Cheng Kung University, Tainan 70101, TAIWAN

*E-mail: ddmorgan@wisc.edu

## Abstract

The effective charge of an element is a parameter characterizing the electromgration effect, which can determine the reliability of interconnection in electronic technologies. In this work, machine learning approaches were employed to model the effective charge ($z^*$) as a linear function of physically meaningful elemental properties. Average 5-fold (leave-out-alloy-group) cross-validation yielded root-mean-square-error divided by whole data set standard deviation (RMSE/σ) values of 0.37 ± 0.01 (0.22 ± 0.18), respectively, and $R^2$ values of 0.86. Extrapolation to $z^*$ of totally new alloys showed limited but potentially useful predictive ability. The model was used in predicting $z^*$ for technologically relevant host-impurity pairs.

**Introduction**

Electromigration (EM) is the biased diffusion of certain species under electric current, and causes significant reliability problems in modern electronic products, *e.g.*, the EM-induced formation of voids/hillocks, which can lead to device malfunction via short circuit and open circuit.[1] Theoretical explanations for the EM effect have included, in chronological order, the semi-ballistic model[2], polarization model[3], back stress model[4], and lattice strain model[5]. Today the driving force of EM is typically considered as a combination of the electron wind force, which is due to electron-ion scattering, and the direct force, which originates from the external electric field. The overall driving force for the EM effect is often formulated as shown in Eq. (1):

$$F = (z_d + z_w)e\rho j = z^* e\rho j \cdots\cdots (1)$$

where $z_d$ is the valence charge associated with the direct force contribution, $z_w$ is the effective charge associated with the electron wind force contribution, $z^*$ is the effective charge sum of the previous two terms of a given species, $e$ is the elementary charge, $\rho$ is the resistivity and $j$ is the current density. The value of $z^*$ is often approximately formulated as shown in Eq. (2):

$$z^* = z_d + z_w \approx z_d + \frac{K}{\rho(T)} \cdots\cdots (2)$$

where $\rho(T)$ is the resistivity of the host, and $K$ is a host- and impurity-specific constant.[2, 6] The species' effective charge sets the scale and diffusion direction of the EM driving force. The larger the effective charge is, the greater the driving force at a given current density level. The negative and positive sign of effective charge represent the diffusion direction toward anode and cathode side, respectively. The techniques of measuring $z^*$ include the marker motion method[2] and radioactive isotope method[7]. Through these techniques, the drift velocity $v$ of a certain species at a given current density and temperature is measured, and the value of $z^*$ is determined following the Einstein relation, as shown in Eq. (3):

$$\frac{v}{j} = \frac{D}{kT} e\rho z^* \cdots\cdots (3)$$

where $D$ is the diffusion coefficient, $k$ is the Boltzmann constant, and $T$ is the absolute temperature. The effective charge is therefore derived from the slope of a $\frac{v}{j}$ vs. $\frac{1}{T}$ plot. The coefficient of variance, *i.e.*, the standard deviation divided by the average value of effective charge, across different research groups measuring different alloy systems may range from *ca.* 4%[8] to 80%[7]. These large variations are due to the fact that the experiments on EM to determine $z^*$ are quite challenging. These experimental challenges consist of the need to conduct measurements at high current density and temperature, the requirement of well-controlled atmospheric conditions, as well as the need to have very long current-stressing times, lasting from a couple of days to months. The large uncertainty existing in certain systems can make developing a reliable database difficult and will introduce significant errors into a machine learning model fit to experimental data. However, results from the above techniques are still considered the state of the art for experimentally determining the effective charge of a given species and useful values are often obtained.

Given the challenges in determining $z^*$, simulating its value is an important and popular topic in EM research. Pioneering quantum mechanical modeling was performed by Bosvieux and Friedel as far back as 1962 using their polarization model. They calculated the $z_w$ as the electrostatic response on a bare ion to electron charge density perturbation due to the presence of defects and scattering under an external electric field.[3] Their seminal modeling approaches have enabled a number of subsequent modeling studies,[6, 9-12] but there is significant controversy about their assumption of weak electron-ion scattering from the Coulomb potential and their method of treating $z_d$.[6, 7] While there has been major progress on the scattering problem, it is still uncertain how to properly model $z_d$.[13] Nevertheless, since the majority contribution to $z^*$ originates from $z_w$, most of the work following Bosvieux and Friedel aimed at improving the simulation of $z_w$. Sorbello in 1973 used first principles calculations based on a polarization model adapted to a pseudo-potential method to replace the weak scattering assumption of Bosvieux and Friedel to calculate $z_w$.[6] The modeling results of Sorbello showed good agreement with experimental measurement for certain alloy systems, such as Na and K, but did not agree well with systems such as Cu, Ag, and Au. It has been suggested this disagreement is likely due to the presence of *d*- or *f*-electrons in noble metals, resulting in different or additional physics governing EM for noble metals.[6, 7] More recently, van Ek and co-workers adapted a Green's function formulation based on the Korringa-Kohn-Rostoker (KKR) method for calculating $z_w$ of impurities in dilute alloy systems.[9-12] The results were improved compared to Sorbello's work, but significant discrepancies still existed in many systems, *e.g.*, Cd, In, Sn, and Sb impurities in a Ag host.[10] Thus, while quantum simulations have made significant progress in modeling $z^*$, they are not yet sufficiently robust for use in generating large databases or even generating data for use in a machine learning models due to potentially large errors versus experiments that are still poorly understood.

Recently, machine learning (ML) methods have been increasingly pursued as a promising materials informatics approach to determine key features controlling a materials property and predicting values outside of those previously measured.[14-22] In this study, a multivariate linear regression (LR) model, which assumes a linear relationship between the input descriptors and the single output quantity, was used for developing a ML model to explore the effective charge of pure elements and impurities in dilute alloy systems. We note that a LR model is a very simple form of ML, but found that more advanced model approaches (specifically, LASSO, Gaussian kernel ridge regression, random forest and decision tree regression) produced similar or worse results at the cost of slower model development and enhanced complexity. A number of statistical analyses were performed for descriptor optimization and assessing the predictive ability of the proposed LR model.

**Methods**

The linear regression was performed with the python library scikit-learn,[23] an open source ML package distributed under BSD license. The model analysis and exploration was primarily performed with the MAterials Simulation Toolkit for Machine Learning (MAST-ML), an open source python package designed to automate machine learning workflows and model assessment.[24] Descriptors consisted of a set of elemental properties constructed using the Materials Agnostic Platform for Informatics and Exploration (MAGPIE) approach (see the SI for additional details of how the MAGPIE approach was implemented in this work). The selection of input descriptors was performed by combining physical intuition and the sequential forward selection algorithm (SFS) as implemented in the open source mlxtend package[25] with the leave-out (LO) alloy-group cross-validation (CV) method and the CV root-mean-square-error (RMSE) as the scoring metric. Here, CV RMSE refers to the RMSE of the left-out (validation) data averaged over all CV splits. A complete literature survey was performed to develop the database and the final data set used in the fitting contains experimental effective charges for 26 dilute alloys and 23 pure metals all measured at, or extrapolated to be at, a homologous temperature (*i.e.*, the temperature of a material as a fraction of its melting point temperature) of $0.9 \pm 0.06$. Statistical analysis, 5-fold CV, LO alloy-group CV and LO-element CV tests were performed to assess the model. The details of the data set, SFS, statistical analysis, and CV methods can be found in the Supplemental Information (SI).

**Data**

To ensure all data used in this paper are easily accessible and adequately archived, we have placed the following files in the SI and on Figshare with DOI 10.6084/m9.figshare.7175072.

1. Figures Data: Fig X.csv and Fig SX.csv contain all the data used to make Figure X and Figure SX in the manuscript and the SI, respectively.

2. Original data sets: The complete databases used in the study, including all effective charges and all descriptors for all the alloy and pure metal systems is titled "Dataset(used)". The complete initially developed database is titled "Dataset(whole)". The text file titled "Reference_dataset" on Figshare lists the references used to obtain the database of effective charges.

**Results and discussion**

An initial set of descriptors (referred to as descriptor set 1) were automatically selected by the SFS algorithm (see SI for the detailed discussion about the descriptor set 1 selection with the SFS algorithm as well as the associated model assessments shown in Fig. S1 and S2, respectively). However, descriptor set one did not seems optimal from a physical perspective, so we modified descriptor set 1 as follows. First, it is well-known that the effective charge is a function of the electrical conductivity and the valence charge of a given species.[6] However, descriptor set 1 included the thermal conductivity instead of the electrical conductivity. This was likely because the thermal conductivity is strongly correlated to the electrical conductivity in metals and the selection process on limited data erroneously picked the less physical quantity. Descriptor set 1 on included *the minimum value of thermal conductivity between the host and impurity*, which we interpreted to be the SFS algorithm attempting to include the physics of the impurity, not the host, as the host thermal conductivity was already in the first descriptor. Therefore, to develop a more physical model, the first two SFS chosen descriptors of (1) *thermal conductivity of host element* and (2) *the minimum value of thermal conductivity between the host and impurity* were manually replaced by (1) *electrical conductivity of host element*, and (2) *electrical conductivity of impurity element,* respectively. We then combined these two descriptors with the next two descriptors in descriptor set 1, which were (3) *the periodic table column difference between the host and impurity*, and (4) *the electronegativity difference between the host and impurity,* froze these four descriptors, and re-ran the SFS to get the new learning curve, as shown in Fig. 1a, and the final descriptor list (note that the difference between host and impurity is taken with the absolute value). The learning curve suggests that the optimal number of descriptors is five (the same as descriptor set 1), since the average RMSE did not significantly decrease when adding more descriptors. The SFS returned the fifth descriptor as (5) *the maximum value of p valence electrons between the host and impurity*. This new descriptor is similar to the previous fifth descriptor in descriptor set 1 (*the compositional average of the number of p valence electrons*), but more reasonable as compositional averages are likely to be poor descriptors in dilute systems. These manually-optimized descriptors will be called descriptor set 2. As shown below, the use of descriptor set 2 yields no significant reduction in the model assessment statistics but provides a much more physically reasonable model.

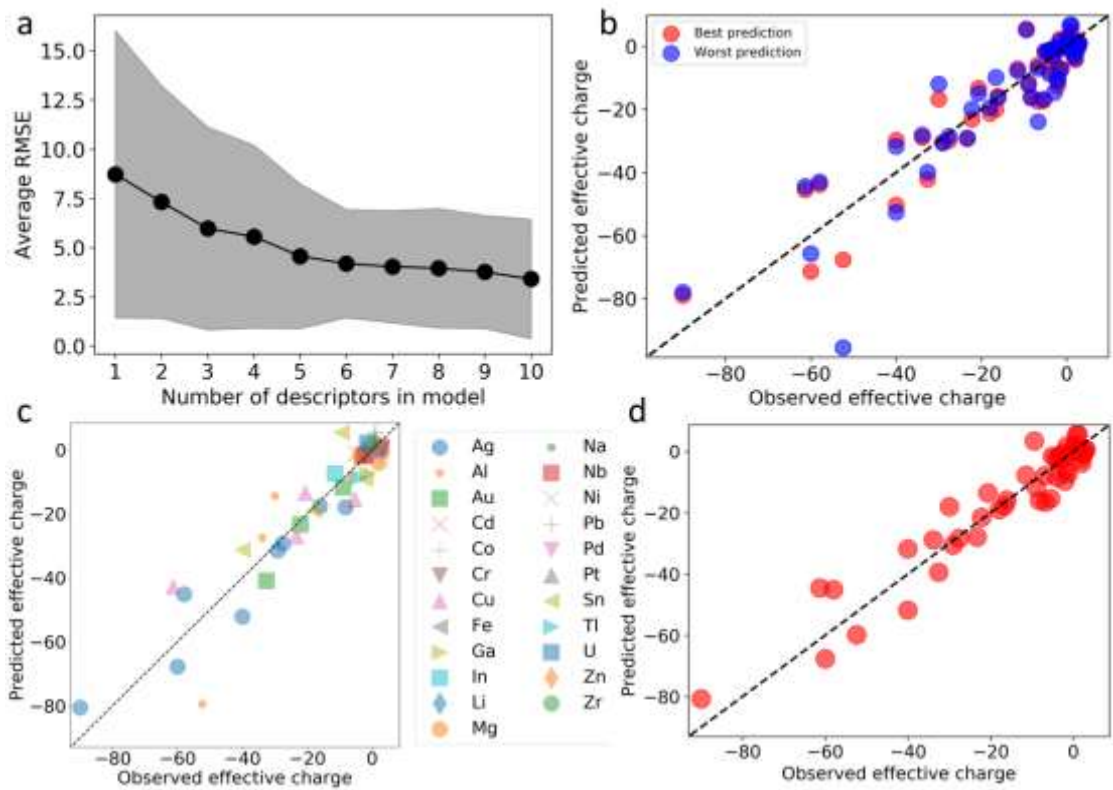

Figure 1. (a) The learning curve with SFS algorithm, (b) the 5-fold CV test (RMSE = 7.84 ± 0.25, RMSE/σ = 0.37 ± 0.01, and $R^2$ = 0.86), (c) the LO alloy-group CV test (RMSE = 4.57 ± 3.67, RMSE/σ = 0.22 ± 0.18, and $R^2$ = 0.86), and (d) the full fit plot (RMSE = 6.21, and $R^2$ = 0.91).

Statistical analysis including the *p*-value test, variance inflation factor (VIF) and Wald test were performed to examine the significance of the manually-optimized (descriptor set 2) descriptors. The data sets were standardized (*i.e.*, to give values of zero mean and unit standard deviation independently for the output z* and for each input descriptor) to obtain the standardized coefficients, and their corresponding standard errors, the *t* statistics, probability value (*p*-value, or *p(>t)*), variance of inflation (VIF) and Wald test *p*-value, are tabulated in Table 1. Details of each statistic can be found in the SI. The low *p*-value indicates that the addition of the five descriptors are statistically significant to the model. The VIF values of approximate one indicates that the descriptors are not strongly linearly dependent and thus multicollinearity is not a concern statistically. The Wald test *p*-value indicates that *the electrical conductivity of host element* has the highest impact on the effective charge, which is consistent with the known physics of this quantity,[2, 6] as shown in Eq. (2). These tests suggest that the five descriptors have a meaningful correlation with the data and that they could be useful in building a ML model for exploring effective charges.

It is helpful both to validate the model and gain physical insight to consider the sign and magnitude of the model parameters. In particular, the present model can be mapped onto the same form as Eq. (2), where the contribution of the intercept and descriptors (2-5) yield $z_d$, and the coefficient of descriptor (1) corresponds to *K* (see SI text and Table S1 for the detailed derivation and discussion). To check if our model is genuinely consistent with the conventional understanding, the $z_d$ and *K* values were calculated with the present ML model and compared with available experimental data (see Table S2 in the SI for the detailed values for comparison). The predicted *K* values are all negative, consistent with experimental values and the general trend of increase in z* for higher host resistance. However, there is no significant correlation between the predicted and experimentally determined values for either $z_d$ or $K$. We made a few attempts to adjust the modeling approach to see if better agreement could be found. First, since $K$ is known as a system-dependent quantity while the present ML model only returned a constant coefficient of descriptor (1), $K$ was further expanded to first order in our descriptors (to give the form shown in (S5) in the SI) to see if the system-dependent characteristics could be better captured. However, expanding $K$ to its the first order did not improve the 5-fold CV RMSE and yielded a poor model. As another approach we noted that, given the form of Eq. (2), it might work better to fit $z^* \times \rho(T) \approx K$, where the approximate equality follows from the fact that most of the contribution of z* often comes from the $\frac{K}{\rho(T)}$ term. However, fitting to this quantity also yielded a quite poor model. This may suggest that the $z_d$ term in Eq. (2) cannot be omitted and has significant importance, that $K$ has a more complex dependence on alloy system than $z^*$, or that there was not enough data to constrain complex models. These aforementioned tests suggest that due to the small data set and the absence of explicit temperature-dependent data for any given system, the ML model is not able to explicitly separate the $z_d$ and $K$ terms. Thus the present ML model cannot be mapped accurately onto Eq. (2), and therefore does not provide predictions for z* at temperatures other than near the homologous temperature of 0.9 ± 0.06. Nevertheless, the present ML model capturing the electrical conductivity of the host element as a descriptor as well as its negative sign is consistent with the previous understanding.

Table 1. A statistical analysis summary of the ML model.

| *Descriptors | Coefficient | Standard error | †$t$ statistics | €$p$ (>$t$) | §VIF | ¢Wald test $p$-value |
|---|---|---|---|---|---|---|
| Electrical conductivity (H) | -0.7827 | 0.049 | -15.974 | $0 \times 10^0$ | 1.15 | N/A |
| Electrical conductivity (I) | 0.3802 | 0.047 | 8.112 | $0 \times 10^0$ | 1.06 | $2.61 \times 10^{-19}$ |
| Column of periodic table (D) | -0.3971 | 0.057 | -6.992 | $0 \times 10^0$ | 1.55 | $3.07 \times 10^{-5}$ |
| Electronegativity (D) | 0.3431 | 0.055 | 6.210 | $0 \times 10^0$ | 1.47 | $1.21 \times 10^{-18}$ |
| # of $p$ valence e⁻ (M) | -0.4225 | 0.058 | -7.284 | $0 \times 10^0$ | 1.62 | $1.31 \times 10^{-6}$ |

* H, I, D and M stand for Host, Impurity, Difference between host and impurity, and Maximum value between host and impurity, respectively. Note that the difference

between host and impurity is taken with the absolute value. The model for the z* is equal to= (-0.7827)×*Electrical conductivity (H)* + (0.3802)×*Electrical conductivity (I)* + (-0.3971)×*Column of periodic table (D)* + (0.3431)×*Electronegativity (D)* + (-0.4225) × *# of p valence e⁻ (M)* + intercept.

† The *t* statistics is the coefficient divided by its standard error.

€ *p(>t)* means the *p*-value associated with the *t* statistics used in testing the null hypothesis which the coefficient is 0.

§ VIF is variance inflation factor, which is defined as $VIF = \frac{1}{1-R_i^2}$, where $R_i$ is the coefficient of multiple determination obtained from regressing a given descriptor $x_i$ on all the other descriptors.

¢ The Wald test is to test the hypothesis that the coefficient of a given descriptor is equal to the *electrical conductivity of host element* and returns the *p*-value.

To check the predictive ability of the model using descriptor set 2, Figs. 1b-d show the parity plot for the best and worst cases of 20 iterations of random 5-fold CV, the LO alloy-group CV, and the full fit to the data set, respectively. The average RMSE of the 5-fold CV shown in Fig. 1b is 7.84 ± 0.25 (see SI for the brief description for the meaning of the reported errors) with the corresponding $R^2$ value of 0.86 (see SI for the details of the $R^2$ calculation), which is slightly better than the previous model made using descriptor set 1 (5-fold CV RMSE is 8.01 ± 0.30 with an $R^2$ value of 0.84) shown in Fig. S2a in the SI. The average RMSE was smaller than the dataset standard deviation ($\sigma$), *i.e.*, 20.96, and yielded an average RMSE/$\sigma$ value of 0.37 ± 0.01 (see SI for the brief description for the meaning of the reported errors), which is significantly less than one and therefore suggests some predictive ability of the model. LO alloy-group CV has RMSE of 4.57 ± 3.67 (RMSE/$\sigma$ of 0.22 ± 0.18) with the $R^2$ value of 0.86 (Fig. 1c), which is also slightly better than the one shown in Fig. S2b in the SI (LO alloy-group CV RMSE is 4.76 ± 3.34 with an $R^2$ value of 0.86). The histogram of residual plot shown in Fig. S3 in the SI shows an approximately normal distribution, further supporting that the choice of the present LR model is appropriate. The RMSE of the full fit shown in Fig. 1d is 6.21 with a full-fit $R^2$ value of 0.91, which is slightly better than when descriptor set 1 was used, as shown in Fig. S2b in the SI (RMSE of 6.97 and $R^2$ of 0.88), again showing that the updated descriptor set 2 is equal or better than those produced by the fully automated SFS (descriptor set 1).

To examine the predictive ability of the model to systems with elements not in the database, a LO element-group CV test was performed, where groups consisted of compositions containing each element were left out (*e.g.*, all compositions containing Ag are one group). The effective charge parity plots for all 23 LO element-group individual tests can be found in Fig. 2. The element name shown in each subplot is the element group left out of the training data, and is thus the validation sub-data set. The groups with only a single data point showed mostly bad prediction, *e.g.*, Co, Ga, Sb, Tl, U and Zr. This was likely in part due to the fact that those element systems were poorly-studied and might have large experimental errors. For example, Ga melts at near room-temperature and it is possible to find a partially liquid phase existing in the sample when performing the electric current experiment, making the *z\** measurement quite uncertain. It is also possible that these less studied systems have some different physics, which could lead to less accurate predictions.

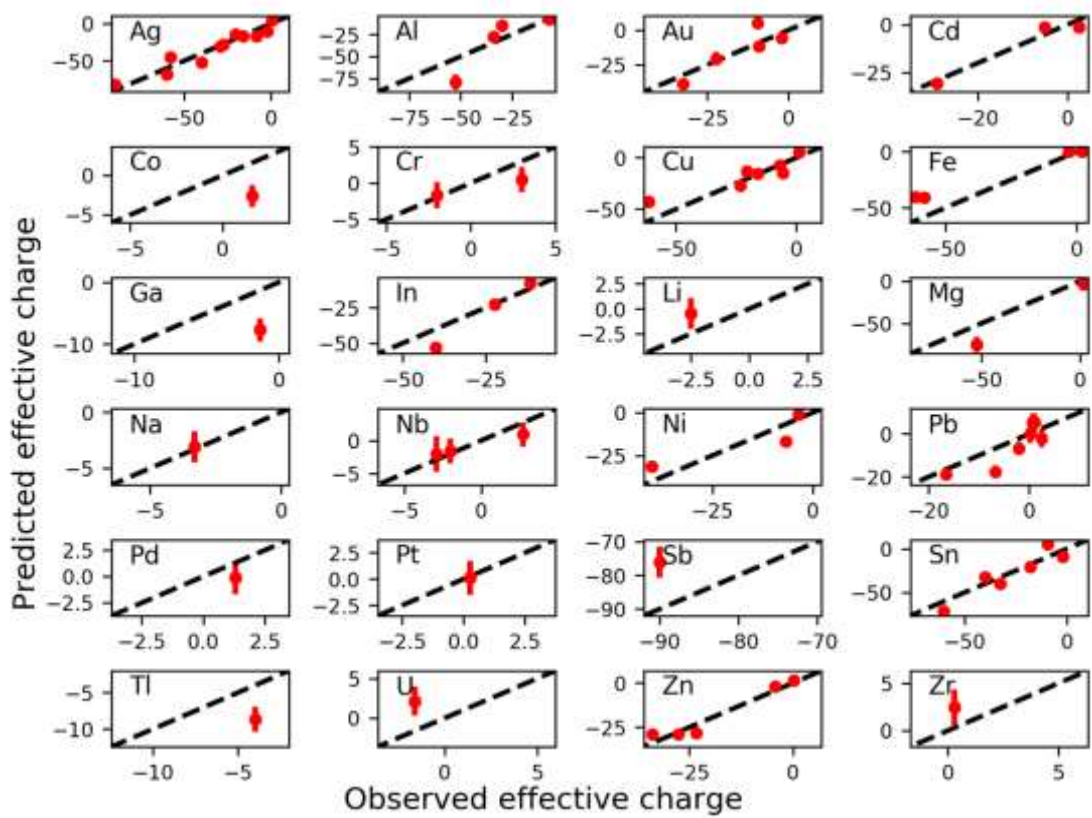

Figure 2. The predicted effective charges from the present LR model versus the experimental ones for leave-out-element group cross-validation test (*i.e.*, for each element group when not included in the training data).

Overall, the LO element-group CV test yielded a RMSE of 6.21 ± 4.76 (RMSE/σ of 0.30 ± 0.23) with a corresponding $R^2$ of 0.89, suggesting some predictive ability for totally new elements not in the training database. Figure 3 further shows the RMSE/σ fitting to the left out validation sub-data set and its corresponding $R^2$ value (note that the σ here is the standard deviation of the validation sub-data set, not the whole data set; see SI for the detailed description for the errors), represented by the filled black bar chart and the filled red circle scattering plot, respectively (see Fig. S4 and Fig. S5 in the SI for the validation sub-data set standard deviation and the average 5-fold CV test for LO element-group model, respectively). The reason only 14 elements are shown in Fig. 3 is because those elements have more than two data points, so the standard deviation can be calculated. All systems show a RMSE/σ value smaller than one and corresponding $R^2$ values higher than 0.70, except for the Cr systems. Even though Cr shows a low $R^2$ value, the qualitative trends, to the extent they can be established with just two points, still appear to be captured. Overall, the LO element-group CV test results are encouraging and suggested some predictive ability for the model on systems with new elements not used in the model training.

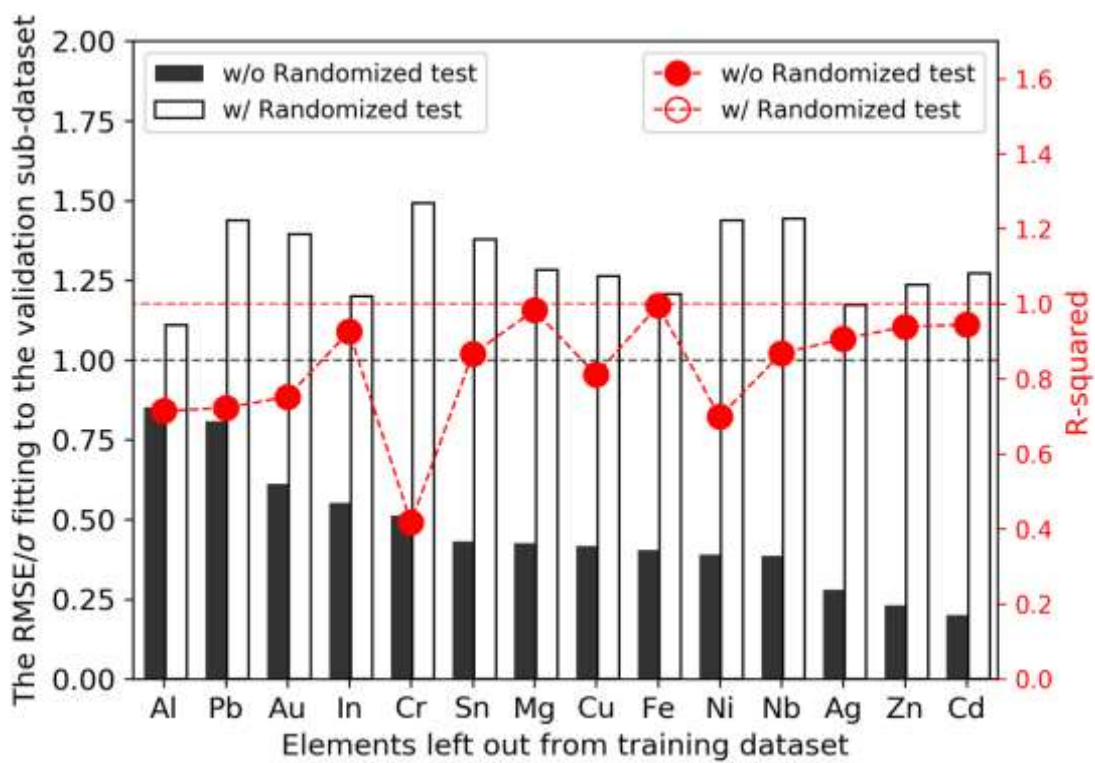

Figure 3. The RMSE/σ and $R$-squared ($R^2$) value for the prediction with and without randomized test.

One concern is that given the small data set with limited sampling of the alloy space, the use of SFS and/or adjustments made during model construction (*e.g.*, choice of LR vs. other methods) might have created apparently physical correlations that are not real. To test if this occurred, we performed what we call a randomized test. A randomized test involves randomly associating each effective charge with a given descriptor, but not the correct one. This gives a new data set which is exactly like the original one in terms of the actual values, but all the physical associations of the descriptors and effective charge have been removed. All the models for the randomized test should be significantly worse than the models for the original data fits if the models for the original data are physically meaningful. The empty black bar chart and the empty red circle scatter plot shown in Fig. 3 are RMSE/σ fitting to the validation sub-data set and its corresponding $R^2$ value when performing the randomized test, respectively (see Fig. S5 in the SI for the average 5-fold CV test for LO element model when performing randomized test). Note that the RMSE/σ and the $R^2$ value are the average values over 20 iterations of the randomized tests and the descriptors are re-selected by SFS in each iteration. All the RMSE/σ values in the randomized test are higher than for the original data (and are all greater than one), and the $R^2$ values for the randomized test are all negative values, indicating that the model built by the randomized test is very poor (see SI text for the $R^2$ calculation details). This result of the randomized test for the fitting process is consistent with there being no correlation or physics in the model. This randomized test result also demonstrates that our approaches did not inadvertently create spurious correlations and further demonstrates the successes of the model for the original data were due to real physical correlations.

Now that we have demonstrated that the model has some physical justification and predictive capability, we attempt to explain the extreme values of $z^*$ in terms of the five descriptors and the intercept, such as Sb in Ag with effective charge of -90.00 or pure Cr with the value of +3.00. Figure 4 shows the contribution of the five descriptors to the associated effective charge value in all systems, as well as the intercept of the model. The contribution means the true value (*i.e.*, not standardized) contributed to the total $z^*$ by each descriptor multiplied by its coefficient and by the intercept. The horizontal axis is the alloy, ordered from the most negative to the most positive measured effective charge. It can be seen that as the effective charge increases, the negative contributions (with negative sign) generally decrease, especially in the *electrical conductivity (H)*, while the positive contributions (with positive sign) appear not to show any decreasing or increasing trend. Overall, these trends suggest that if the host is a good conductor and the impurity is not, with a small difference of the electronegativity between the impurity and the host, the effective charge of the impurity is expected to be negative. The periodic table column difference and the number of $p$ valence electrons only makes the effective charge even more negative. For instance, Sb shows a very negative effective charge in a Ag host due to a large difference in periodic column between Sb and Ag (*Column of periodic table (D)*), as well as the larger number of $p$ valence electrons of Sb (*# of p valence e$^-$ (M)*). On the contrary, pure elements like pure Cr show a positive value of $z^*$ mainly due to a net small negative contribution of the electrical conductivity (*i.e.*, the coefficient of *Electrical conductivity (I)* is smaller than *Electrical conductivity (H)*) and zero contribution from the two difference descriptors, which are cancelled by the positive contribution of the intercept. Further, the intercept is the only term left that can lead to a positive value in pure element systems with zero contribution for the *# of p valence e$^-$ (M)*, *e.g.*, Cr, Nb, Fe, Mg, Co, Pd, Zr and Pt. As discussed above, the intercept is likely to be part of the $z_d$ value in these pure element cases based on Eq. (2). The intercept plays a role in the pure elements which show positive $z^*$.

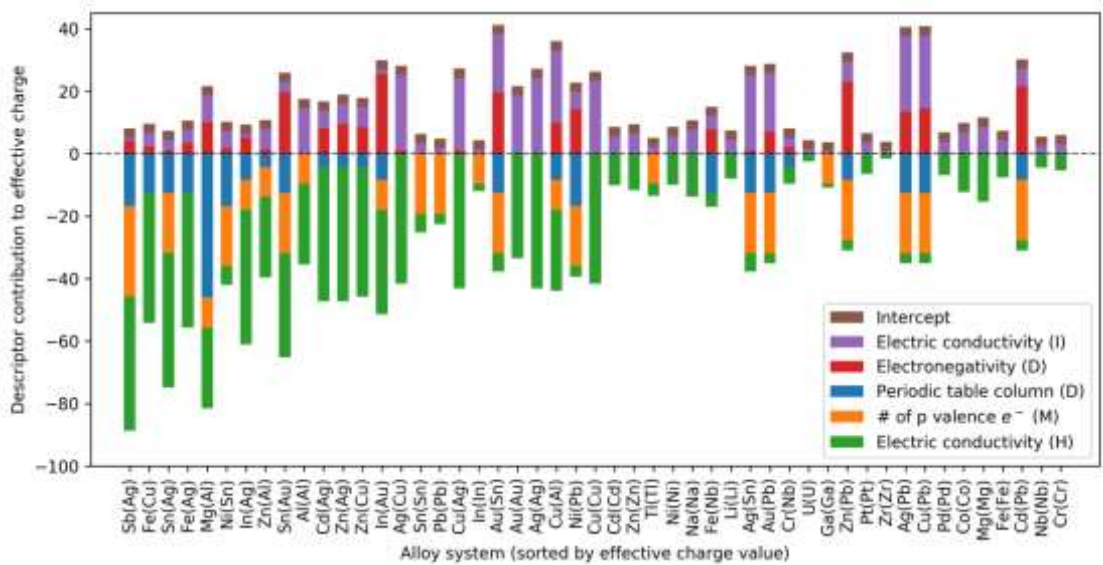

Figure 4. The contribution of each descriptor to the effective charge value shown in a stacked bar chart. Elements shown in parentheses on the horizontal axis represent the host, while the ones outside the parentheses are the impurity. The H, I, D and M symbol stand for Host, Impurity, Difference between host and impurity, and Maximum value between host and impurity, respectively.

Figure 5 shows predictions for the effective charge of impurities across the periodic table in 6 different hosts: Al, Ag, Au, Co, Cu, and Sn. These hosts were chosen because they are commonly-used alloy systems in electronic products. The exact predictions are given in Table S3 in the SI and available on Figshare (see data section). The bright teal and the bright pink colors in the color bars represent the negative and positive extremes of the effective charge, respectively. The predictions were assessed by comparing with the existing literature where possible. These comparisons are necessarily of a qualitative nature as any cases with relevant quantitative measurements have been included in the database already. The comparisons are as follows: (1) the EM velocity of Cu in Ag was found to be insignificant relative to the EM velocity of Ag in Ag.[26] Our prediction shows that the effective charge of Cu in Ag relative to the Ag host is *ca.* 0, consistent with this experimental result. (2) Pd was found to migrate slower than Sn in a Cu host under electric current.[27, 28] Our prediction shows that the effective charge of Pd in Cu is smaller than Sn in Cu, again consistent with the experiments. (3) The effective charge of Sb in Au was found to be larger than Au in Au .[29] Our prediction also agrees with this trend. (4) The effective charge of Au in Sn was found to be larger than Ag in Sn,[8] but our prediction shows the opposite relation. The origin of this disagreement is not clear at this time, but it is not outside the expected ranges of $2\sigma = 5.92$ error. Since most of the predictions are new and there are no associated experimental observations, the validation is very limited at this time.

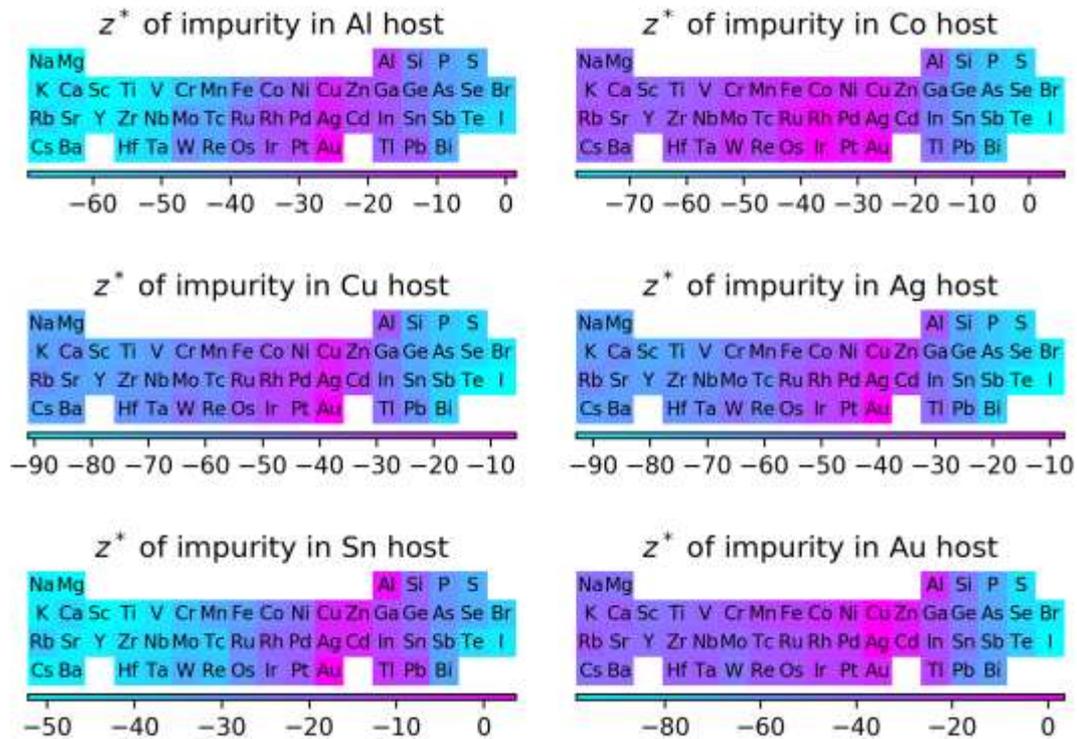

Figure 5. The exploration of effective charge of impurities in Al, Co, Cu, Ag, Sn, or Au host.

It is worth noting that Co is proposed to be a new alternative to Cu in the back-end-of-line (BEOL) interconnection material for the latest technology node, which is essentially the physical size of a transistor made in a particular technology, so obtaining an improved understanding of EM in this host is particularly important as very little data is available.[30] Our predictions suggest that Ag, Au, Cu, Ir, Pt, Pd, Rh, and Ru in Co will have $z^*$ values close to zero, leading to little EM effect for those elements. However, elements such as Al, Bi, Mg, P, Sb, Si and Sn in Co may lead to significant EM effect.

These predictions may help the design of the next-generation interconnection materials in the electronic industry.

Even though some discrepancies exist in the present prediction and the experimental findings, with the average model 5-fold CV RMSE (RMSE/σ) of 7.84 ± 0.25 (0.37 ± 0.01) with $R^2$ value of 0.86, LO alloy-group CV RMSE (RMSE/σ) of 4.57 ± 3.67 (0.22 ± 0.18) with $R^2$ value of 0.86, the robust statistical deviation from zero of the fitting coefficients, the consistency between the underlying physics of the descriptors and classical EM models, and the reasonable success of the LO element-group test, we are confident that the present LR model is able to give qualitative guidance for unknown host-impurity systems. A major limitation of the present ML model is that it is likely only valid for binary dilute alloy systems and pure metals at the homologous temperature of 0.9 ± 0.06, as this is the type of data that was used in the fitting. However, the modeling approach used here is very flexible and the model could be easily improved and extending to more complex systems by adding new data as it becomes available.

**Conclusion**

In this study, a machine learning linear regression model was developed to explore the effective charge ($z^*$) for electromigration of impurities in binary dilute alloy systems and pure metals at the homologous temperature of $0.9 \pm 0.06$. The most effective descriptors included (1) *the electrical conductivity of the host element*, (2) *the electrical conductivity of the impurity element*, (3) *the periodic table column difference between the host and impurity*, (4) *the electronegativity difference between the host and impurity* and (5) *the maximum value of p valence electrons between the host and impurity*, and were selected by a combination of a sequential forward selection algorithm and manual selection based on domain-specific knowledge of the physics governing $z^*$. Standard statistical analyses including the *p*-value, variance inflation factor, and Wald test show that these five descriptors made a significant contribution to the model, that multicollinearity of descriptors is not an issue, and that the most important descriptor is the electrical conductivity of the host, respectively. 20 iterations of 5-fold CV, leave-out alloy-group CV, and leave-out element-group CV yielded average values as follows: 5-fold CV - RMSE/σ = $0.37 \pm 0.01$, $R^2 = 0.86$, leave-out alloy-group CV - RMSE/σ = $0.22 \pm 0.18$, $R^2 = 0.86$, and leave-out element-group CV - RMSE/σ = $0.30 \pm 0.23$, $R^2 = 0.89$, together indicating some significant predictive ability of the present model. A leave-out element-group test and a randomized test suggest the predictive ability to unknown systems, and ensures the present fitting has physical meaning, respectively. The descriptor list suggests that if the host is a good conductor and the impurity is not, with a small difference of the electronegativity between the impurity and the host, the effective charge of the impurity is expected to be a negative value. The periodic table column difference and the number of $p$ valence electrons of the impurity makes the effective charge value more negative. The descriptors provided new information for the understanding of the origin of effective charge. Predictions of the effective charges of impurities across the periodic table within 6 often-used hosts including Al, Ag, Au, Co, Cu and Sn were made with the present model. A semi-quantitative model is obtained in the present work and the approach can be easily applied to develop improved models as new data becomes available in the future. The present machine learning model can potentially be utilized to accelerate the design of materials used in electrical interconnections and other applications where EM may play a role.

## Acknowledgement

The authors Yu-chen Liu and Shih-kang Lin gratefully acknowledge the financial support from the Ministry of Science and Technology (MOST) in Taiwan (106-2628-E-006-002-MY3) and Graduate Student Study Abroad Program (GSSAP) (107-2917-I-006-008). Benjamin Afflerbach, Ryan Jacobs and Dane Morgan gratefully acknowledges support from the National Science Foundation (NSF) Software Infrastructure for Sustained Innovation (SI2) award No. 1148011.